\documentclass{emulateapj}

\newcommand{\lsun}{\mbox{L$_\odot$}}
\newcommand{\lint}{\mbox{L$_{int}$}}
\newcommand{\msun}{\mbox{M$_\odot$}}

\newcommand{\mean}[1]{\mbox{$\langle#1\rangle$}} 
\newcommand{\av}{\mbox{$A_V$}} 

\newcommand{\lbol}{\mbox{$L_{bol}$}} 
\newcommand{\tbol}{\mbox{$T_{bol}$}} 

\newcommand{\mdisk}{\mbox{$M_{disk}$}} 
\newcommand{\menv}{\mbox{$M_{env}$}} %
\newcommand{\rdisk}{\mbox{$R_{disk}$}} %
\shorttitle{Disks in Class 0 Protostars}
\shortauthors{Enoch et al.}

\begin{document}

\title{Disk and Envelope Structure in Class 0 Protostars: II. High Resolution Millimeter Mapping of the Serpens Sample}

\author{Melissa L. Enoch (1), Stuartt Corder (2), Gaspard Duch\^{e}ne (1,3), Douglas C. Bock (4), Alberto D. Bolatto (5), Thomas L. Culverhouse (6), Woojin Kwon (7), James W. Lamb (6), Erik M. Leitch (6,8), Daniel P. Marrone (8,9), Stephen J. Muchovej (6), Laura M. P\'{e}rez (10), Stephen L. Scott (6), Peter J. Teuben (5), Melvyn C. H. Wright (1), and B. Ashley Zauderer (5,11)}

\affil{
(1) Department of Astronomy, University of California at Berkeley, 601 Campbell Hall, Berkeley, CA, 94720, USA \\
(2) NRAO/ALMA-JAO, Av. Apoquindo 3650, Piso 18, Las Condes, Santiago, Chile \\
(3) UJF-Grenoble 1 / CNRS-INSU, Institut de Plan\'etologie et d’Astrophysique de Grenoble (IPAG) UMR 5274, Grenoble, F-38041, France
(4) Combined Array for Research in Millimeter-wave Astronomy, Big Pine, CA 93513, USA  \\
(5) Department of Astronomy and Laboratory for Millimeter-wave Astronomy, University of Maryland, College Park, MD 20742, USA \\
(6) Owens Valley Radio Observatory, California Institute of Technology, Big Pine, CA 93513, USA \\
(7) Department of Astronomy, University of Illinois, Urbana, IL 61801, USA \\
(8) Department of Astronomy and Astrophysics, University of Chicago, 5640 S. Ellis Ave. Chicago, IL 60637, USA \\
(9) Hubble Fellow \\
(10) Department of Astronomy, California Institute of Technology, 1200 East California Blvd, Pasadena, CA 91125, USA \\
(11) Harvard-Smithsonian Center for Astrophysics, 60 Garden Street, Cambridge, MA 02138, USA 
}

\begin{abstract}

We present high-resolution CARMA 230~GHz continuum imaging of nine deeply embedded protostars in the Serpens Molecular Cloud, including six of the nine known Class 0 protostars in Serpens.
This work is part of a program to characterize disk and envelope properties for a complete sample of Class 0 protostars in nearby low-mass star forming regions.  
Here we present CARMA maps and visibility amplitudes as a function of $uv$-distance for the Serpens sample.  Observations are made in the B, C, D, and E antenna configurations, with B configuration observations utilizing the CARMA Paired Antenna Calibration System.
Combining data from multiple configurations provides excellent $uv$-coverage ($4-500$~k$\lambda$), allowing us to trace spatial scales from $10^2$ to $10^4$~AU.  
We find evidence for compact disk components in all of the observed Class~0 protostars, suggesting that disks form at very early times ($t<0.2$~Myr) in Serpens.  
We make a first estimate of disk masses using the flux at 50~k$\lambda$, where the contribution from the envelope should be negligible, assuming an unresolved disk.  The resulting disk masses range from $0.04~\msun$ to $1.7~\msun$, with a mean of approximately 0.2~\msun.
Our high resolution maps are also sensitive to binary or multiple sources with separations $\gtrsim250$~AU, but significant evidence of multiplicity on scales $<2000$~AU is seen in only one source.

\end{abstract}

\section{Introduction}

Circum-stellar and circum-protostellar disks are an important component of the star formation process, transferring mass from the surrounding envelope to the growing central protostar.  While disk formation is a natural result of collapse in a rotating core, it is still not known how soon after protostellar formation the disk appears, or how massive it is at early times.  Simple analytical theories of star formation suggest that centrifugally supported disks should start out  quite small (radius $<10$~AU), and consequently with very low mas, and grow with time \citep*{tsc84}.  
More recent simulations of self-gravitating and viscous disks indicate that relatively massive disks can form at very early times \citep[e.g.][]{vorobyov09}.
Magnetically supported disks can also be much larger (radii up to 1000~AU; \citealt{gs93}), and thus more massive at early times.

To understand the formation and growth of disks, and their relationship to the accretion process, it is critical to observe and characterize disks in embedded sources, where most of the mass of the growing star is being accreted.  
Directly observing disks in the embedded phases is much more difficult than in more evolved pre-main sequence stars, as they are hidden within the dense, extincting protostellar envelopes.

\input{tab1}

Embedded protostars are defined here as protostars with substantial ($>0.1~\msun$) 
envelope masses remaining.  We will also refer to young sources by class, where the bolometric temperature \tbol\footnote{The bolometric temperature is the temperature of a black-body with the same mean frequency as the observed spectral energy distribution ($\tbol=1.25\times10^{-11} \mean{\nu}$~K; \citealt{ml93}) and is a better measure of the evolutionary state than the infrared spectral index \citep{enoch09a}.} \citep{ml93,chen95} is used to divide sources into Class~0 ($\tbol \le 70$~K), Class~I (70 K$< \tbol \le 650$~K), and Class~II (650 K$<\tbol \le 2800$ K).
Here we assume that these classes correspond to an evolutionary sequence \citep[e.g.][]{rob06,andre94}: in Class~0 the protostar has accreted less than half its final mass ($M_{*} < \menv$), in Class~I more than half of the mass has been accreted ($M_* > \menv$), and in Class~II the envelope has dispersed, leaving only a circum-stellar disk.  

In the last two decades a number of studies have focused on detecting disks in embedded protostars \citep[e.g.][]{chan95,brown00,loon00,harv03,jorg05b,eisner05,aw07}.  
These studies have found that disks are well established in Class I, with typical masses of $0.01-0.5~\msun$.
Although circum-protostellar disks have been detected in a number of Class 0 sources, most detailed studies of Class~0 sources have been limited to the brightest or most well-known objects due to instrumental limitations, a lack of complete target samples, and the difficulty of separating disk and envelope emission in such young sources.

\citet{jorg07,jorg09} have recently completed a large Submillimeter Array (SMA) survey of ten Class 0 and ten Class I sources, finding similar disk masses in Class 0 and Class I.  Although \citet{jorg09} see no clear evolutionary signature in disk flux or mass, they do see an increase in the ratio of disk mass to envelope mass with evolutionary stage.
The SMA study still relies on well known sources from different regions, however, rather than a complete or well defined sample.  A study characterizing ``typical'' Class~0 disks, in a  sample that is complete down to some well defined luminosity or envelope mass limit, is still needed.

We have recently begun a program to characterize 
disks and envelopes in a complete sample of Class 0 protostars in nearby low mass star-forming regions.  Combining CARMA 230~GHz continuum imaging and \textit{Spitzer} IRS mid-infrared (MIR) with radiative transfer modeling of this sample will help to address several fundamental questions about the structure and evolution of the youngest protostars: 1) How soon after the initial collapse of the parent core does a circum-protostellar disk form? 2) What fraction of the total circum-protostellar mass resides in the disk, and does this fraction vary with time?  3) Are large ``holes'' in the inner envelope \citep[e.g.][]{jorg05} common at early times?

Millimeter maps and MIR spectra provide complementary approaches to these questions. The amount of flux escaping at $\lambda \lesssim 50~\micron$ from embedded protostars is sensitive to the opacity close to the protostar, and thus the envelope structure \citep[e.g.][]{jorg05}.  While the MIR flux is insensitive to disk properties, high resolution millimeter continuum mapping directly detects emission from dust grains in the disk.  Millimeter observations with good $uv$-coverage and calibration, combined with radiative transfer models, can separate the disk from the envelope and constrain the disk mass and size.

Our ultimate goal is to characterize the disk mass, size, and inner envelope structure of typical low-mass Class~0 protostars, and to quantify any trends with evolutionary indicators.
In \citet{enoch09b}, hereafter Paper I, we presented results for Serpens FIRS~1, a well known Class 0 source, as a pilot for the full program.  We found that disks can be quite massive at very early times; radiative transfer modeling (RADMC; \citealt{dd04}) compared to CARMA visibilities and spectral information indicated that FIRS~1 has a disk with mass $\mdisk=1.0 ~\msun$ and radius $\rdisk=300$~AU, making up approximately 13\% of the total circum-protostellar mass in a source that is about $10^5$~yr old.

Here we present high-resolution millimeter imaging of 9 embedded protostars in the Serpens Molecular Cloud, including 6 of the 9 known Class 0 sources in the cloud.  
The Combined Array for Research in Millimeter-Wave Astronomy (CARMA\footnote{http://www.mmarray.org/}) is an ideal instrument to study the disk properties of deeply embedded protostars.  With 15 antennas operating at $\lambda=1$~mm, CARMA provides the excellent $uv$-coverage required to separate disk emission from envelope emission in young protostars.

In Section~\ref{samplesec} we describe the Serpens sample, and in Section~\ref{obssec} outline the CARMA observations and data reduction.  CARMA maps and visibilities are presented in Section~\ref{mapsec}, followed by an initial estimate of disk masses from the visibility amplitude at large $uv$-distances in Section~\ref{disksec}.  Individual sources are discussed in Section~\ref{indsec}.  We comment on the observed multiplicity fraction in our sample in Section~\ref{multsec}.

\section{Serpens Sample}\label{samplesec}

Recent large surveys of nearby molecular clouds at mid-infrared and (sub)millimeter wavelengths have made it possible to define complete samples of Class 0 protostars based on luminosity or envelope mass limits \citep[e.g.][]{hatch07,jorg08,dun08,enoch09a,evans09}.
Our study is based on the complete sample of 39 Class 0 protostars in the Serpens, Perseus, and Ophiuchus molecular clouds, identified by \citet{enoch09a}, identified by comparing large-scale \textit{Spitzer} IRAC and MIPS with Bolocam  \citep{glenn03} 1.1~mm continuum surveys of the three clouds.
Here we focus on the Serpens cloud, which contains 34 embedded protostars, 9 Class~0 sources and 25 Class~I, and is complete to envelope masses $\menv\gtrsim0.25~\msun$ and internal luminosities $\lint\gtrsim0.05~\lsun$.

The Serpens molecular cloud is an active star forming region with high extinction and a high density of young stellar objects (YSOs).   It has been studied extensively at near-infrared, far-infrared, submillimeter, and millimeter wavelengths \citep[e.g.][]{ec92, hb96, lars00, davis99, casali93, ts98}.
We adopt a distance of $d=415\pm25$~pc  based recent VLBA trigonometric parallax measurements of a young stellar object in Serpens  \citep{dzib10}.  Note that this is significantly larger than the $d=260$~pc that we have assumed in our previous work (Paper I; \citealt{enoch07}; \citealt{enoch09a}).
To compare to our previous disk and envelope masses in Serpens, masses given here should be decreased by a factor of 2.5 and physical sizes by 1.6.  Mass ratios will remain unchanged. 

Targets are listed in Table~\ref{targettab}, including the ``Ser-emb\#'' name from \citet{enoch09a}, \textit{Spitzer} position, bolometric luminosity (\lbol), bolometric temperature (\tbol), and total envelope mass (\menv).  
Our goal here is to image all known Class 0 protostars in Serpens with high resolution and high fidelity with CARMA.  
\citet{enoch09a} found 9 Class 0 protostars in Serpens.  Three of the 9 known Class~0 source were not included in the 230~GHz observations, however.
These three sources were either not detected, or only marginally detected, in preliminary 110~GHz lower-resolution CARMA maps (see Section~\ref{appendsec}).

All undetected sources are in confused regions, with multiple Spitzer sources associated with a single Bolocam 1.1~mm core;
at least one source in each core is detected and included here.  
The undetected sources do not have particularly low luminosity or low envelope mass (see Table~\ref{targettab}), so excluding them does not bias the survey by luminosity or circum-protostellar mass.
In addition to six Class~0 targets, we include three Class~I sources that are near the Class~0/I cutoff.

Two sources in our sample (Ser-emb~4 and emb~6) were observed as part of the  \citet{hoger99} interferometric study of four deeply embedded protostars in Serpens, but of the two only Ser-emb~6 was detected.  The other two \citet{hoger99} sources were classified as Class~I by \textit{Spitzer}. 

\begin{figure}[!ht]
\hspace{-0.7in}
\includegraphics[width=4.75in]{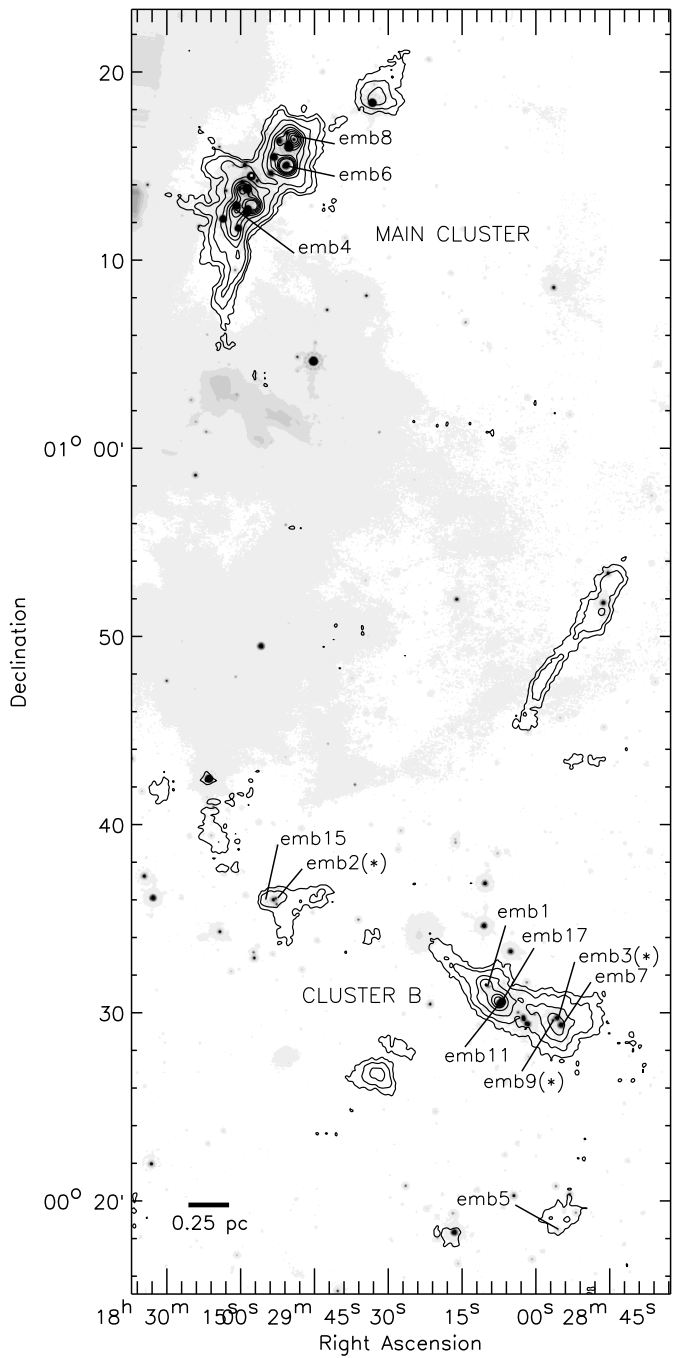}
\caption
{Overview of the Serpens Main cluster and Cluster B region, with a finding chart for embedded protostars observed with CARMA.  The \textit{Spitzer} $24~\micron$ map from c2d is shown in grayscale, with Bolocam 1.1~mm continuum contours overlaid (contours are $50,100,200,400,600...1400$~mJy beam$^{-1}$).  Although the Serpens cloud, as defined by the $\av=6$ contour, extends beyond the region shown, all embedded protostars are found within this area.  ``emb\#'' designations are from \citet{enoch09a}, and sources marked with a ``(*)'' are Class~0 protostars not included in the 230~GHz study (see Sections~\ref{samplesec} and \ref{appendsec}).
\label{genfig}}
\vspace{0.1in}
\end{figure}

\begin{figure*}[!ht]
\includegraphics[width=7.5in]{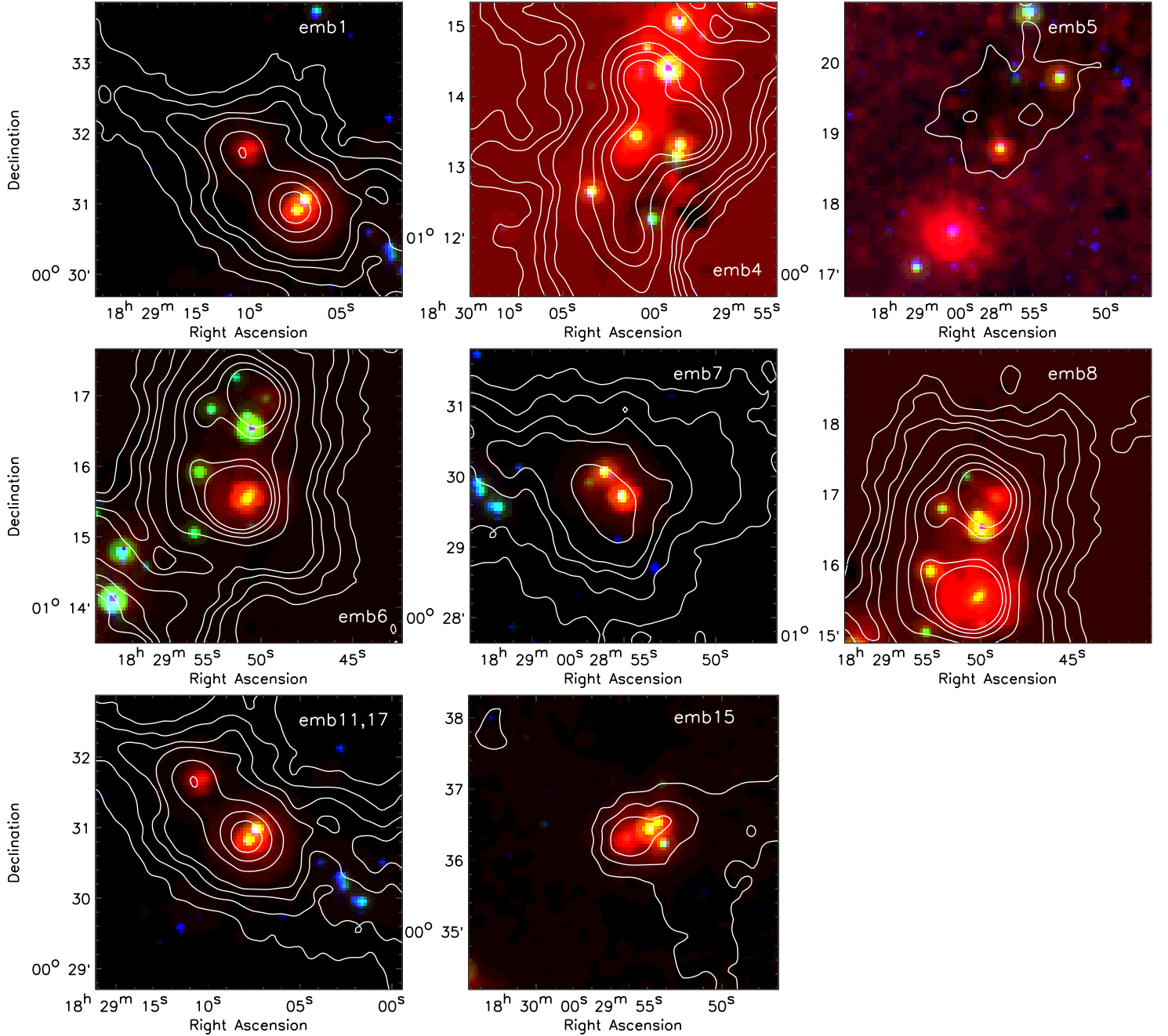}
\vspace{0.2in}
\caption{
\textit{Spitzer} three color ($8~\micron, 24~\micron, 70~\micron$) images of the 9 embedded protostars imaged here with CARMA.  Bolocam 1.1~mm continuum contours are shown at $50,100,150,200,400,600,800,1000$~mJy beam$^{-1}$.  Each target is centered; Ser-emb~11 and emb~17 are $<15\arcsec$ apart and are shown in the same panel.
\label{spitzfig}}
\vspace{0.1in}
\end{figure*}

As described in \citet{enoch09a}, \lbol\ and \tbol\ are calculated from the spectral energy distribution (SED) from $1.25~\micron$ to $1100~\micron$.  Broadband infrared data for our targets are taken from the ``From Molecular Cloud Cores to Planet Forming Disks'' \textit{Spitzer} Legacy program (``Cores to Disks'' or c2d; \citealt{evans03}), which imaged  one square degreee in the cloud with IRAC and MIPS \citep{harv06,harv07}.  The same region was mapped at $\lambda=1.1$~mm with the Bolocam bolometer array at the Caltech Submillimeter Observatory \citep{enoch07}.
Together with public 2MASS catalogs, these data provide wavelength coverage at 1.25, 1.65, 2.17~$\micron$ (2MASS), 3.6, 4.5, 5.8, 8.0~$\micron$ (IRAC), 24, 70, 160~$\micron$ (MIPS), and 1100~$\micron$ (Bolocam).  

The envelope mass \menv\ is calculated from the total $\lambda=1.1$~mm single dish flux, assuming the envelope is optically thin at 1.1~mm, a dust opacity of $\kappa_{1mm} = 0.0114$~cm$^2$~g$^{-1}$ \citep{oh94}, a dust temperature of $T_D=15$~K, and $d=415$~pc.  Here we treat the envelope mass as the total circum-protostellar mass, as it may include some flux contribution from a disk component.

Figure~\ref{genfig} gives an overview of the Serpens environment with \textit{Spitzer} $24~\micron$ (grayscale) and Bolocam 1.1~mm continuum (contours) maps.  The locations of our embedded protostar targets are indicated.  Most are associated with the Serpens Main cluster to the north or  Cluster B \citep{djup06,harv06} to the south, but a few are more isolated.
Three color \textit{Spitzer} images ($8, 24, 70~\micron$) of the 9 targets imaged with CARMA are shown in Figure~\ref{spitzfig}, with Bolocam 1.1~mm contours as in Figure~\ref{genfig}.  SEDs of each source from $1.25~\micron$ to $1100~\micron$ can be found in \citet{enoch09a}.

\input{tab2}

\section{Observations and Data Reduction}\label{obssec}

Continuum observations at $\nu=230$~GHz ($\lambda=1.3$~mm) were completed with CARMA, a 23 element interferometer consisting of nine 6.1-m, six 10.4-m, and eight 3.5-m antennas.
The 6.1-m and 10.4-m antennas were used to obtain 230~GHz continuum observations in the B ($100-1000$~m baselines), C ($30-350$~m), D ($11-150$~m), and E ($8-66$~m) antenna configurations between October 2007 and January 2010.  
Three to four sources were observed in each track; although the $uv$-coverage is not identical for each target, every source was observed in each configuration, and at least $\pm3$~hours around transit.  
Integration times varied based on the expected flux from single dish observations. 

Three correlator bands were configured for continuum observations with 468.75 MHz bandwidth, for a total bandwidth of 2.82~GHz (two sidebands per band).  
A bright quasar (1751+096) was observed approximately every $15-20$ minutes, to be used for complex gain calibration.  For B configuration tracks, a secondary calibrator (1830+063) was also observed every 15 minutes.
Absolute flux calibration was accomplished using 5 minute observations of Uranus, Neptune, or MWC~349.  The overall calibration uncertainty is approximately $\pm20$\%, from the reproducibility of the phase calibrator flux on nearby days. 
A passband calibrator, typically 3C454.3, was observed for 15 minutes during each set of observations, and either optical \citep{corder10} or radio pointing was performed every one to three hours.  

Data from different array configurations were combined to provide excellent $uv$-coverage from $4$ to $500~ k\lambda$, providing sensitivity to physical scales from $170$~AU to $2\times10^4$~AU.  
Small 7-point mosaics were made in the compact configurations (D and E) in order to mitigate spatial filtering by the interferometer.  Mosaicing provides better sensitivity to the spatially extended protostellar envelopes.

Observations in the most extended B configuration utilized the CARMA Paired Antenna Calibration System (C-PACS) to correct for phase variations on sub-minute timescales \citep{perez10}.  
For the longest B configuration baselines, C-PACS pairs  6.1-m and 10.4-m antennas (science array) with 3.5-m antennas (reference array) that operate at $\lambda=1$~cm.  The reference array continually monitors the atmospheric phase using a strong point source close to the science target (here 1751+096, also the gain calibrator, which is $\sim13\deg$ away), while the science array slews between the science targets and gain calibrator.  The delay measured by the reference array is later used to correct the science data for atmospheric variations (see \citealt{perez10} for details).  
The 3.5-m antennas were tuned to 30~GHz, with a 14-band wideband correlator providing 7~GHz of bandwidth.  

Calibration and imaging are accomplished with the MIRIAD data reduction package \citep*{sault95}.  
Calibration of each track consists of applying line-length calibration, passband correction, phase correction on $15-30$~min intervals, amplitude calibration on 60~min intervals, and absolute flux calibration.

For B configuration data, delays are calculated from the 30~GHz observations of 1751+096 and applied to each paired baseline on 4~s intervals after passband calibration.  
The 230~GHz gain calibrator observations are then used to apply longer-interval (15~min) phase corrections, to remove slowly varying instrumental delay differences between the reference and science arrays.
C-PACS corrections usually work best for close ($<6\deg$) atmospheric calibrators (Zauderer et al., in preparation); given that 1751+096 is $13\deg$ from our science targets, we only apply C-PACS calibration for tracks where it significantly improves phase coherence on the secondary calibrator (1830+063, $5\deg$ from the science targets). When applied, C-PACS calibration improves the signal to noise of the peak flux of 1830+063 by a factor of $1.5-6$.

Images are formed by inverting the calibrated visibilities using multi-frequency synthesis with natural weighting to optimize signal to noise, cleaned with a Steer CLEAN algorithm \citep*{steer84}, and restored with a Gaussian fit to the synthesized beam.  Beam sizes for the combined maps (including all configurations) are approximately $1\arcsec$, corresponding to a physical resolution of 415~AU.

We also calculate the azimuthally averaged visibility amplitude as a function of $uv$-distance.  
Due to mosaicing in the compact arrays, visibilities must be combined over multiple pointing centers, and are calculated in three steps.  The visibilities are first modified in phase to account for any offset between the source position and the pointing position.  The visibilities are then divided by the primary beam response at the location of the peak of the flux in the various pointings.  Finally, we plot the weighted mean of the visibilities over the different pointing positions.  With this method, very extended components or bright multiple sources can introduce errors in the visibility amplitudes.  However, if the emission is very compact relative to the $30\arcsec$ primary beam, the error introduced by this scaling of the visibilities from the offset pointings is minimal.

\section{Maps and visibilities}\label{mapsec}

\begin{figure*}
\begin{center}
\includegraphics[width=6.8in]{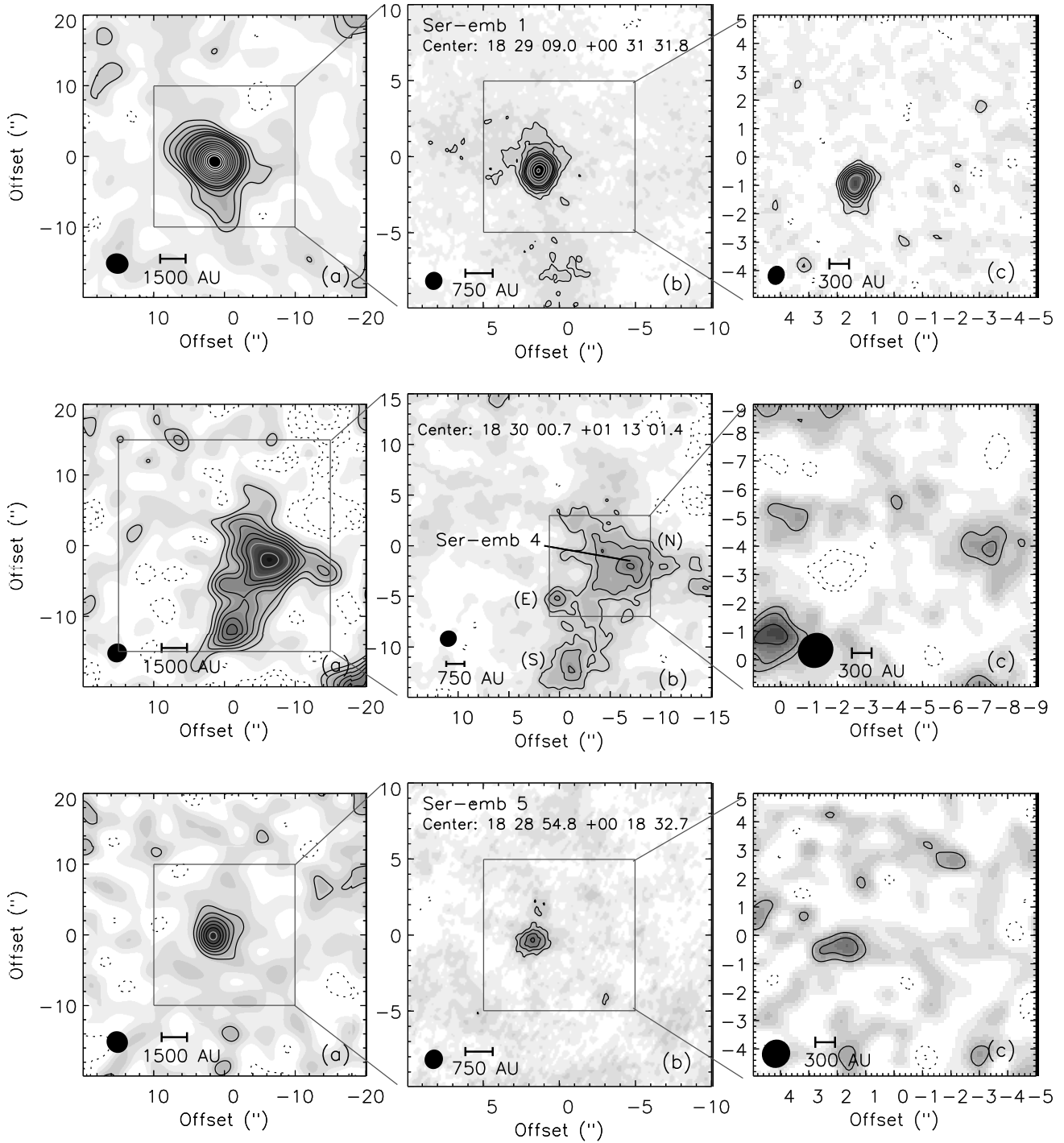}
\caption
{CARMA 230~GHz images of Serpens deeply embedded protostars.  The combined maps, shown in panel (b), include data from B, C, D, and E configurations, with 7-point mosaics in D and E.  
Large-scale and small-scale emission maps are shown in panels (a) and (c), respectively; maps in panel (a) are made including only $uv$-distances $<50$~k$\lambda$, and maps in panel (c) with $uv$-distances $>50$~k$\lambda$ (with the exception of Ser-emb~4, which includes $uv$-distances $>30$~k$\lambda$ due to low signal-to-noise on long baselines).
Contours are shown at $3,5,..15,20...60,80...140~\sigma$, where $\sigma$ is given in Table~\ref{carmatab}, with lighter contours starting at $15~\sigma$ and negative contours indicated by dotted lines.
Synthesized beam sizes (approximately $1\arcsec$ or 400~AU for combined maps) are indicated by solid ellipses. 
Offsets from the pointing centers are likely due to errors in the \textit{Spitzer} positions, which are determined by the shorter wavelength data where outflows can cause extended or offset emission.
\label{mapfig}}
\end{center}
\end{figure*}

The resulting 230~GHz maps are shown in Figure~\ref{mapfig}.  In addition to maps including all data, two maps including only $uv$-distances $<50$~k$\lambda$ and $>50$~k$\lambda$ are shown, to better illustrate any extended and compact emission, respectively. 

The 230~GHz peak flux, total flux, and deconvolved size of each detected source, determined from a Gaussian fit (MIRIAD's imfit) to the combined map, are given in Table~\ref{carmatab}.  The rms noise level in the central region, excluding sources, is also listed (ranging between $0.9-8$~mJy beam$^{-1}$), along with the synthesized beam size, or resolution (typically $\sim1\arcsec$ or 415~AU). Multiple entries are given when more than one source is detected in the CARMA map.

Plots of the visibility amplitude versus $uv$-distance for each embedded protostar are shown in Figure~\ref{visfig}.
Extended envelope structures, as expected for embedded Class~0 protostars, are visible as an amplitude peak at $uv$-distances $<20~k\lambda$ in the visibility plots (Figure~\ref{visfig}).  Only Ser-emb~15 does not show evidence for a strong peak at $uv$-distance $<20$~k$\lambda$.
The peak visibility amplitude is equivalent to the total 230~GHz flux of the source\footnote{Note that this can be larger than the total flux listed in Table~\ref{carmatab} because a Gaussian fit does not always capture the full extent of an extended source.}.
The interferometer does filter out flux at $uv$-distances less than $4~ k\lambda$ (physical scales $>2\times10^4$~AU), corresponding to the separation of the closest antenna pairs, so the total flux observed by CARMA will often be less than the total single dish flux. 

A point source, such as an unresolved disk, will appear as a constant amplitude in the visibility plot, while a resolved disk will contribute an additional wide Gaussian component to the visibilities.  
Although the most common interpretation of a compact visibility component is a disk, without dynamical information or independent constraints on the envelope structure, compact emission could also be attributed to a power law envelope \citep{loon03} or non-axisymmetric envelope structure \citep{loon07}.  With these caveats in mind, here we interpret compact emission as evidence of a circum-protostellar disk.

We can see immediate evidence for a compact disk component, as non-zero amplitudes at $uv$-distances $>50~k\lambda$, in most sources (e.g. Ser-emb~4, emb~7,emb~11, emb~17, emb~15).  Other sources have significant, but non-constant, flux for $uv$-distance $50-100~k\lambda$, which may be due to a partially resolved disk (e.g. Ser-emb~1, emb~6, emb~8).

\begin{figure*}
\begin{center}
\includegraphics[width=6.8in]{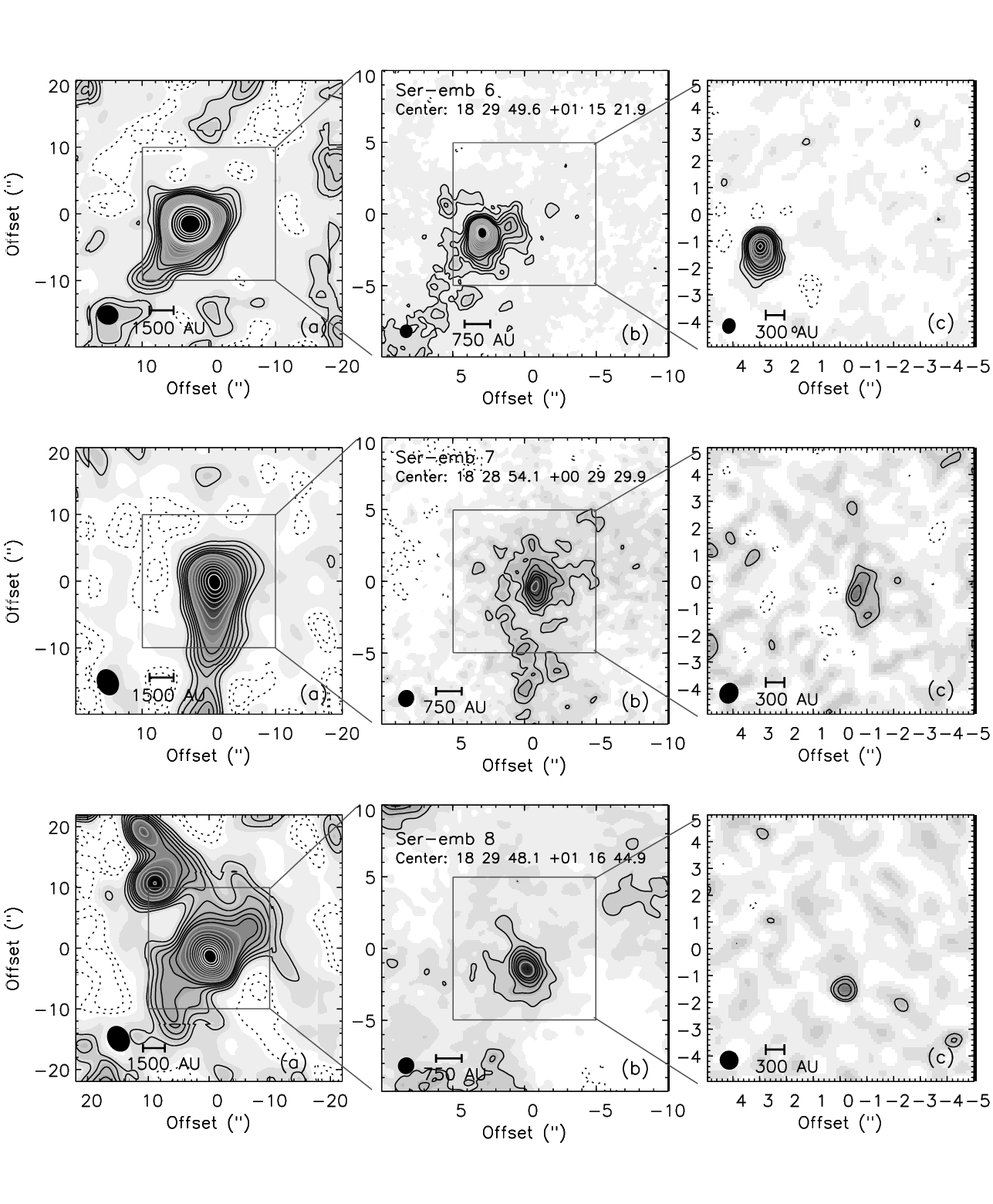}
\caption
{Figure 3 cont.}
\end{center}
\end{figure*}

\begin{figure*}
\begin{center}
\includegraphics[width=6.8in]{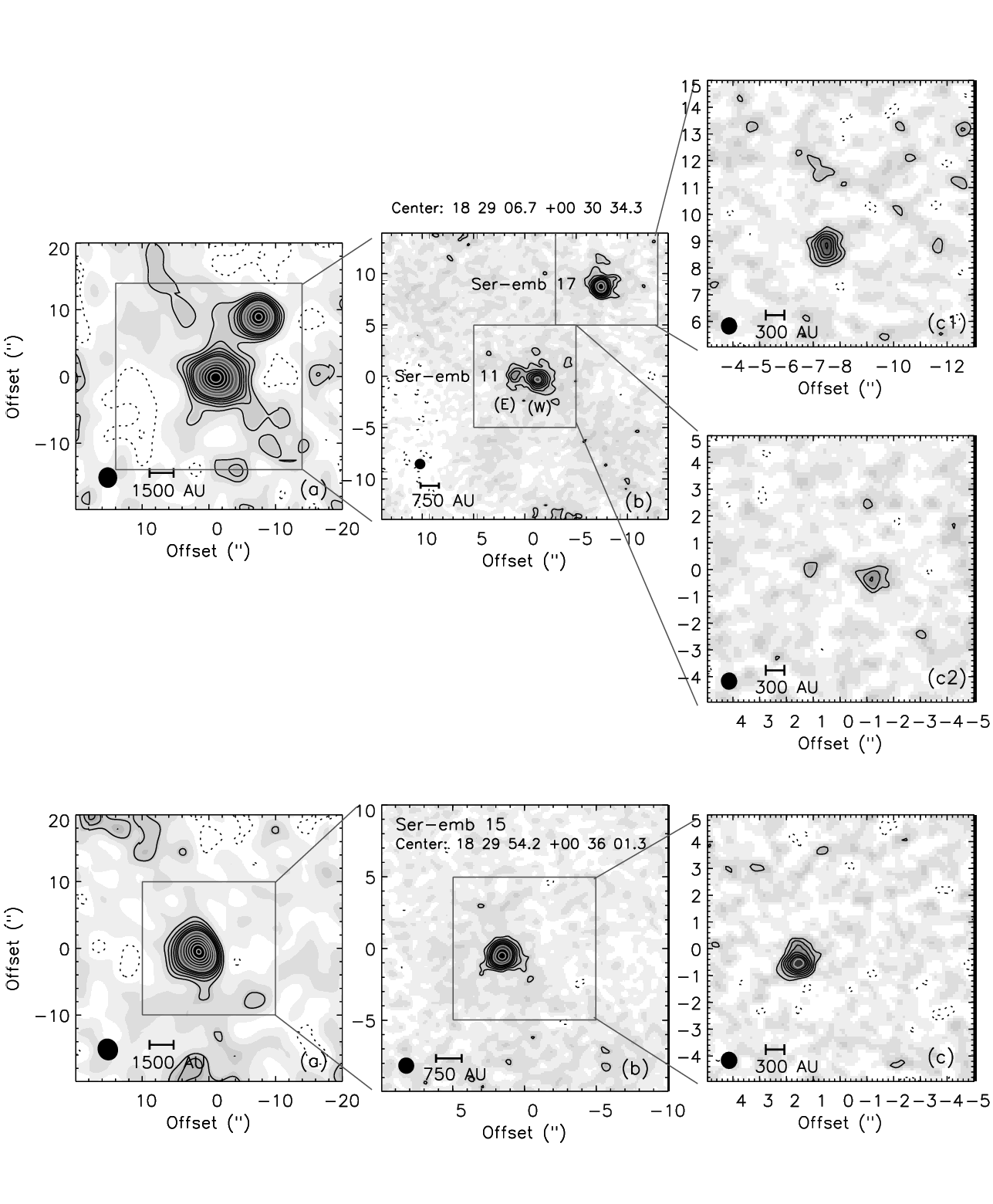}
\caption
{Figure 3 cont.}
\end{center}
\end{figure*}

\begin{figure*}
\includegraphics[width=6.5in]{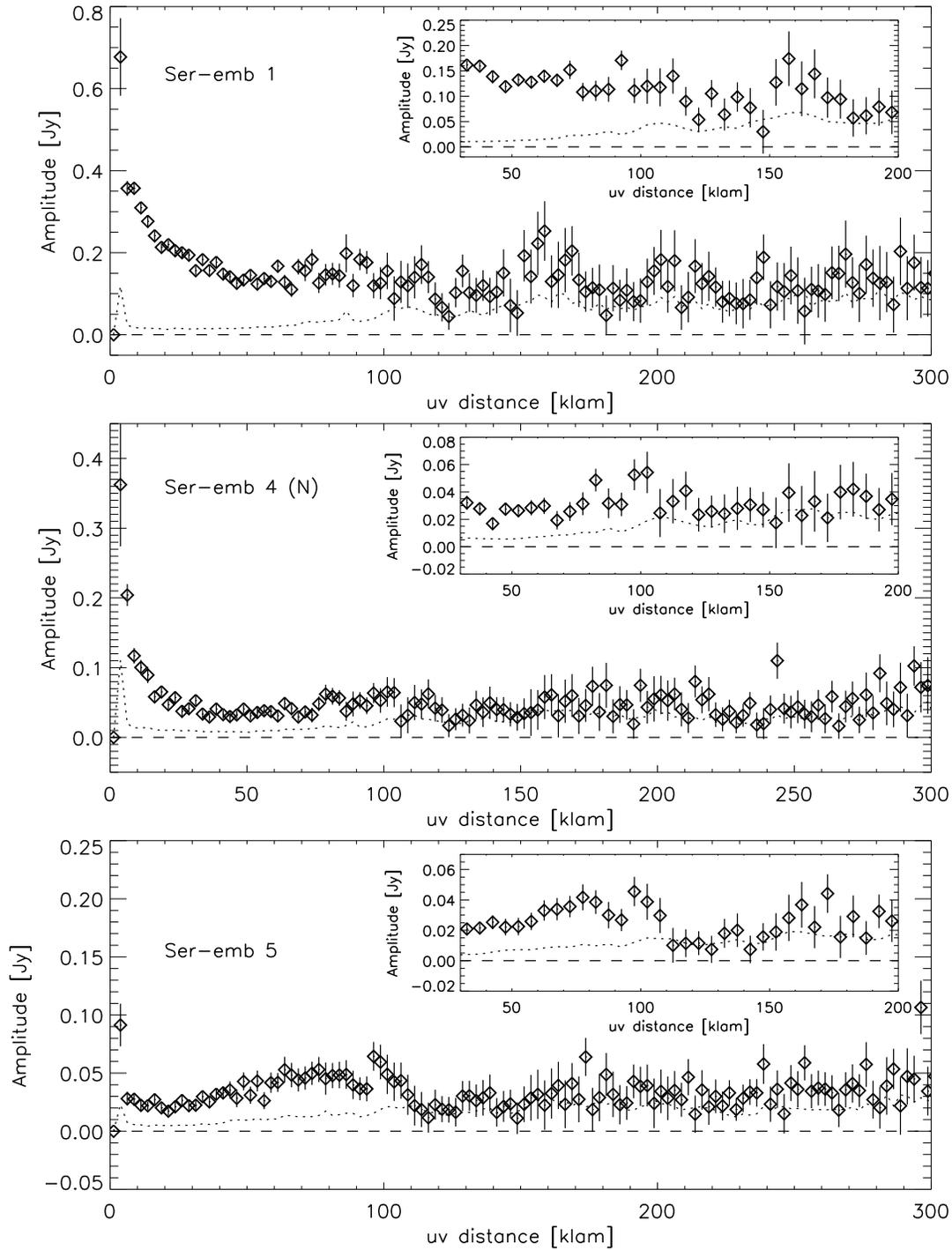}
\caption
{Azimuthally averaged visibility amplitudes as a function of $uv$-distance for the Serpens embedded protostars.  The value expected for zero signal, or bias, is shown for reference (dotted line). Averaged visibilities are vector averaged at the measured source position (Table~\ref{carmatab}), and not the pointing center.  Error bars reflect the variable $uv$-coverage, and atmospheric decorrelation on the longest baselines ($>100$ or $>200$~k$\lambda$ depending on source flux); they do not include the $\sim20$\% absolute calibration uncertainty.  Visibilities are binned in $2.5$~k$\lambda$ increments; insets show the region where disk emission is expected, with $5$~k$\lambda$ bins to increase signal to noise.
\label{visfig}}
\end{figure*}

\begin{figure*}
\includegraphics[width=6.5in]{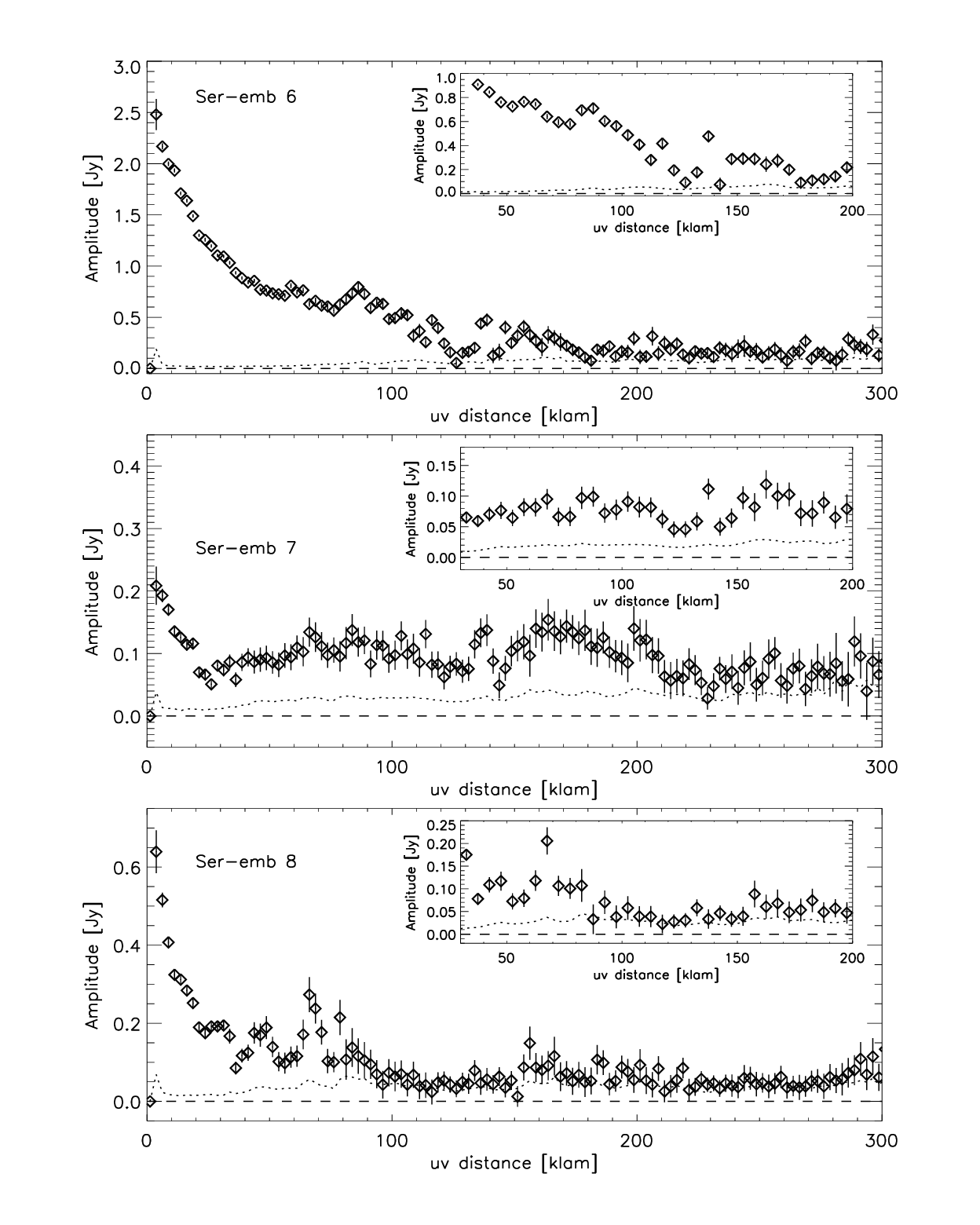}
\caption
{Figure~\ref{visfig} cont.}
\end{figure*}

\begin{figure*}
\includegraphics[width=6.3in]{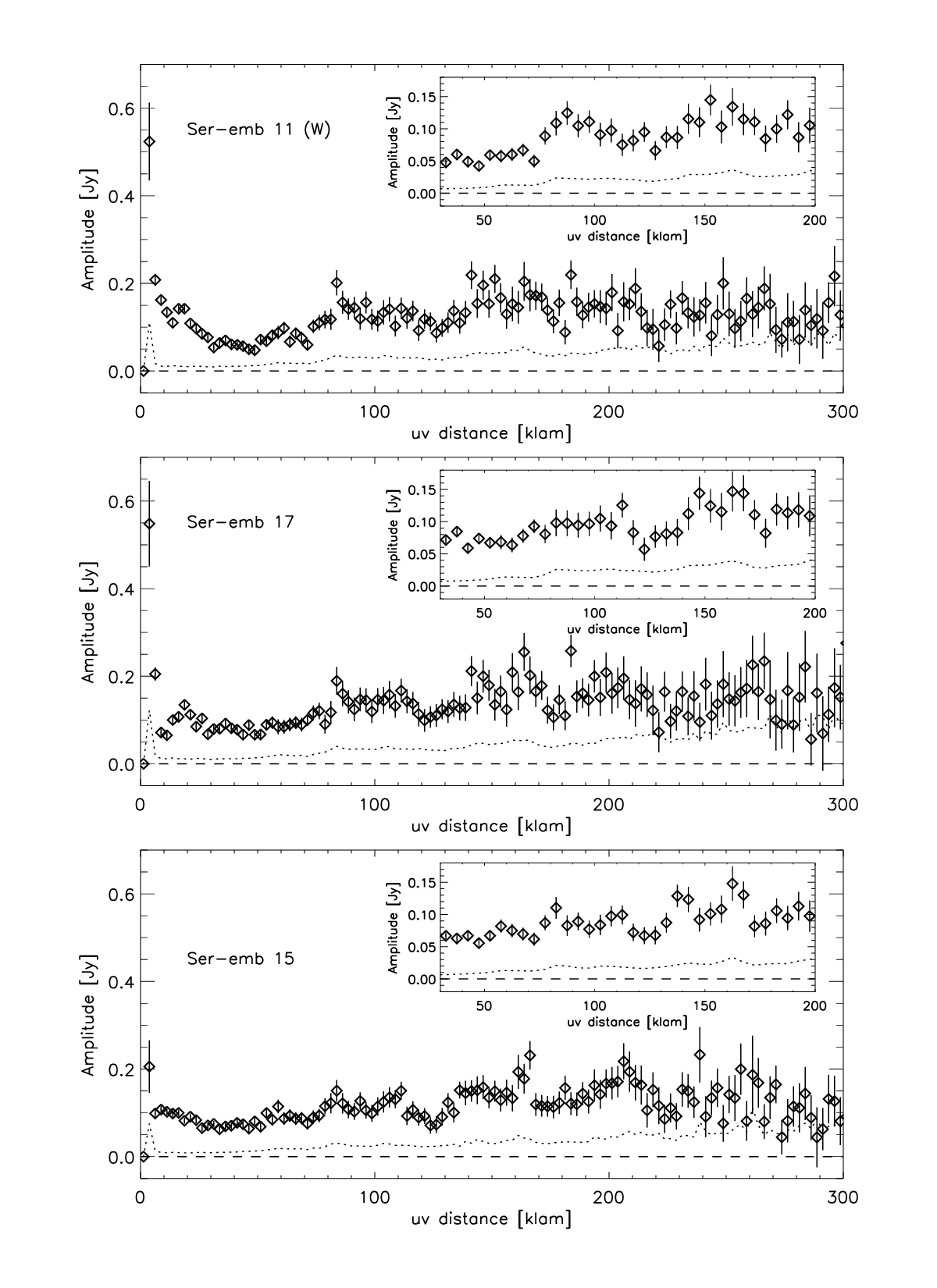}
\caption
{Figure~\ref{visfig} cont.}
\end{figure*}

\section{Estimated disk masses}\label{disksec}

Accurately measuring disk masses requires comparing the 230~GHz visibilities and spectral information with detailed radiative transfer models, as in Paper~I.  Radiative transfer models such as RADMC \citep{dd04} simultaneously model the envelope and the disk, taking into account any envelope component at small scales, as well as a possibly resolved disk.  In addition, radiative transfer models self-consistently solve for the temperature as a function of position.  Work to constrain the disk and envelope structure of these sources with grids of radiative transfer models is currently ongoing (Enoch et al., in prep.).

For now, we make a first estimate of the disk mass using the 230~GHz flux at a $uv$-distance of $50$~k$\lambda$.
An unresolved disk will have a constant flux  independent of $uv$-distance, while the resolved envelope flux falls off quickly with $uv$-distance.   Radiative transfer models of interferometric observations of a number of embedded protostars have indicated that protostellar envelopes contribute little emission at baselines longer than $50$~k$\lambda$ \citep[e.g.][]{km90,harv03,jorg05}.
Using similar models to separate the disk and envelope in Ser-emb~6, we find that the envelope flux for $uv$-distances $>30$~k$\lambda$ is nearly zero (see Figure~6 of Paper~I), contributing $<2$\% of the total flux at $50$~k$\lambda$.

The choice of $50$~k$\lambda$, as was adopted by \citet{jorg09}, is a good compromise between minimizing the envelope contribution and maximizing the signal-to-noise of the disk, which falls off with increasing baseline length (due to decreased $uv$ coverage).

To calculate the mass, we assume that the disk is optically thin at $\nu=230$~GHz:
\begin{equation}
\mdisk = \frac{d^2 S_{50k\lambda}}{B_{1.3mm}(T_D) \kappa_{1.3mm}}, \label{masseq}
\end{equation}
where $S_{50~k\lambda}$ is the flux at 50~k$\lambda$, $d=415$~pc is the cloud distance, and $B_{1.3mm}$ is the Planck function for a dust temperature of $T_D=30$~K \citep[e.g.][]{visser09,jorg09}.  
We assume the dust composition is similar in the disk and envelope, taking the dust opacity per gram of gas at $\lambda=1.3$~mm, $\kappa_{1.3mm} =0.009$~cm$^2$~g$^{-1}$, from Table~1 column~6 of \citet{oh94}, for dust grains with thin ice mantles.

Here we assume that there is a negligible contribution by the envelope to the flux at $50$~k$\lambda$.  If approximately 30\% of the flux at $50$~k$\lambda$ can be contributed to the envelope, as found by \citet{jorg09} based on radiative transfer models of 8000~AU envelopes, disk masses could decrease by 30\%.

On the other hand, if any of the disks are partially resolved (i.e. have radii larger than 170~AU), the flux at $50$~k$\lambda$ will underestimate the disk mass.  For Ser-emb~6, the flux at $50$~k$\lambda$ suggests a mass that is 30\% smaller than the mass from detailed radiative transfer modeling \citep{enoch09b}.
Furthermore, if any part of the disk is optically thick, or the outer regions of the disk have a dust temperature well below $30$~K, the true mass of the disk could be considerably higher than our estimates.  
We therefore assign uncertainties to our estimated disk masses of at least $\pm50$\%.

\begin{figure}[!hb]
\hspace{-0.15in}
\includegraphics[width=3.6in]{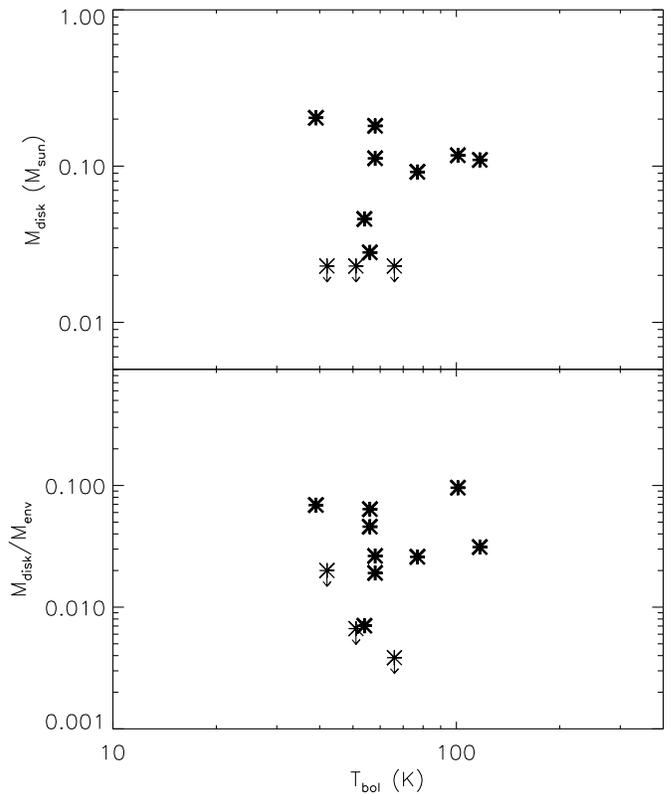}
\vspace{-0.4in}
\caption{
Estimated disk masses (top) and ratio of the estimated disk mass to envelope mass (bottom), as a function of bolometric temperature.  
Upper limits for three Class~0 sources in our sample are based on preliminary 110~GHz maps (Section~\ref{appendsec}).  No strong trend with \tbol\ is seen for either the disk mass or disk to envelope mass ratio.
\label{mdiskfig} }
\end{figure}

Estimated disk masses are given in Table~\ref{masstab}, and are plotted as a function of bolometric temperature, a measure of the evolutionary state, in Figure~\ref{mdiskfig}.  
We do not see a trend with evolutionary state in the narrow $\tbol$ range observed.
The mean disk mass in our 9 source sample is $0.33~\msun$.  

\input{tab3}

We must also account for the three Class~0 sources not included here due to non-detections in preliminary 110~GHz maps.   
If Ser-emb~2, emb~3, or emb~9 had a compact disk with $\mdisk \gtrsim0.02~\msun$ they would have been detected at 3mm (see Section~\ref{appendsec}), so we assume that these sources do not have significant disks.  
Thus, we find evidence for circum-protostellar disks in 6 of the 9 Class~0 sources in Serpens.  If we include these non-detections, we find a mean disk mass $<\mdisk>=0.2~\msun$.

The disk to envelope mass ratio (\mdisk/\menv) is also shown versus \tbol\ in Figure~\ref{mdiskfig}, where \menv\ is from Table~\ref{samplesec}.  Disk masses are between 1\% and 10\% as large as envelope masses in our sample.  The mean \mdisk/\menv\ is 6\%.  Again we see no clear trend in the disk to envelope mass ratio with \tbol\ within this young sample.  

Our high detection rate in this complete sample suggests that disks must form very early in the star formation process.  The Class~0 phase likely lasts approximately 0.2~Myr \citep{enoch09a}, and we have detected disks in 6 of the 9 Class 0 targets, which represent all Class~0 sources with $\lint\gtrsim 0.05\lsun$ and $\menv \gtrsim 0.25\menv$ in the approximately $3.5\deg$ cloud area.  Thus our results suggest that 2/3 of embedded protostars in Serpens have formed an $\mdisk>0.02~\msun$ disk within the first 0.2~Myr. 

Although there is no overlap between the two samples, the disk masses for our complete Serpens sample are roughly consistent with the \citet{jorg09} Class~0 disk masses.  We also see similar disk to envelope mass ratios in Class~0, ranging from $1-10$\% in both samples.  

\citet{jorg09} found average disk masses in both Class~0 and Class~I of $\mdisk=0.05~\msun$, while \citet{aw07} found a lower average Class~I disk mass in Ophiuchus ($0.015~\msun$).  Our mean Class~0 disk mass is somewhat higher, $0.2~\msun$, but this may be a result of temperature variations with luminosity \citep[e.g.][]{jorg09}, or the fact that we have not corrected for the envelope contribution at 50~k$\lambda$.
Given the uncertainties, our results appear to be consistent with the conclusion of \citet{jorg09}, that disk masses are fairly constant through the embedded phases of protostellar evolution.

Such a scenario is supported by recent hydrodynamics simulations by \citet{vorobyov09}, which find time-averaged disk masses in Class 0 and Class I of $\sim 0.1~\msun$ for both viscous and self-gravitational disks.
More robust measurements of disk masses and similar studies of the Class~0 populations in other clouds are needed to confirm these results.  

\section{Individual Sources}\label{indsec}

\subsection{Ser-emb 1}
Ser-emb~1 is located in Serpens Cluster B (Figure~\ref{genfig}), and is associated with the single dish Bolocam 1.1~mm core Ser-Bolo~15 \citep{enoch07}.  Ser-emb~1 was identified as a Class~0 source by \citep{djup06}.  
Ser-emb~1 has the lowest bolometric temperature (39~K), and thus presumably is the least evolved, of all the protostars in Serpens.  It is relatively isolated; a single \textit{Spitzer} mid-infrared source is associated with the approximately $10^4$~AU diameter Bolocam core, and the nearest protostar is $\sim70\arcsec$ ($3\times10^4$~AU) away.

The 230~GHz map of Ser-emb~1 (Figure~\ref{mapfig}) shows a single compact source, which is centered on the mid-infrared source.  The total flux from a Gaussian fit is 215~mJy (Table~\ref{carmatab}), but the peak in the visibility plot (Figure~\ref{visfig}, right) indicates a total flux in the CARMA data of 560~mJy.  
After scaling the 270~GHz ($\lambda=1.1$~mm) Bolocam flux to 230~GHz, assuming a dust spectral index of $\beta=1.5$, we find that CARMA recovers approximately 95\% of the single dish flux.

The visibility plot indicates substantial flux at intermediate $uv$-distances ($30-100$~k$\lambda$), which could be explained by either a resolved disk or a very compact envelope.  The disk mass estimated from the flux at $50$~k$\lambda$ is $0.28~\msun$, or 9\% of the total circum-protostellar mass, although this may be severely underestimated if the disk is partially resolved. 

\subsection{Ser-emb~4}
Ser-emb~4 is located in the Serpens Main cluster, in a crowded region (several other protostars within $1\arcmin$).  It is associated with the Bolocam core Ser-Bolo~28, as are Ser-emb~19 and Ser-emb~29.  Most of the single dish flux, and thus most of the envelope mass, is likely associated with Ser-emb~4, which has a much lower \tbol\ (54~K compared to 129~K and 374~K, respectively).

The  230~GHz map reveals three distinct sources; the northernmost is associated with the Ser-emb~4 mid-infrared source, while the other two appear to have no counterpart at $\lambda<70~\micron$.  The eastern and southern sources are located $10\arcsec$ (4200~AU) and $15\arcsec$ (6200~AU) from the primary northern source, respectively.

Ser-emb~4 (N) has a diffuse extended envelope that is partially resolved out by the small synthesized beam.  The extended nature of the envelope is also apparent in the narrow peak in the visibilities at small $uv$-distances.  The peak visibility amplitude is 250~mJy, only 19\% of the total single dish flux.  This difference is accounted for by the fact that the Bolocam core breaks into several sources with the higher resolution of CARMA.
A constant amplitude of $\sim25$~mJy is seen out to $uv$-distance $\sim 100$~k$\lambda$, suggesting an unresolved disk component with $\mdisk \sim 0.05~\msun$, approximately 1\% of the total circum-protostellar mass.  
The two additional sources in the CARMA map also likely contribute to the single dish flux, making this 1\% fraction a lower limit.

\subsection{Ser-emb~5}
Ser-emb~5 is located south of Cluster B, and is one of the few embedded protostars not in one of the two main clusters.  It is associated with the extended Bolocam core Ser-Bolo~7, as is Ser-emb~25, but the two sources are separated by $>1\arcmin$.  Ser-emb~5 has a low bolometric luminosity ($\lbol=0.4~\lsun$) and low envelope mass ($\menv=0.6~\msun$).  It is the only Class~0 source in Serpens that is considered a Very Low Luminosity Object (VeLLO; see \citealt{dun08}).  VeLLOs are believed to be either very low mass protostars (i.e. proto-brown dwarfs), or stellar mass protostars with very low mass accretion rates \citep{dun08}.

The 230~GHz map contains a single, relatively compact source.  
The peak visibility amplitude is 68~mJy, accounting for 60\% of the single dish flux.
The flux at 50~k$\lambda$ suggests a disk mass of $0.04~\msun$, or 6\% of the total circum-protostellar mass.  Ser-emb~5 has the lowest disk mass of our nine 230~GHz targets. 

\subsection{Ser-emb~6}
Ser-emb~6 is a well known Class~0 protostar also known as Serpens FIRS~1 \citep{hwj84} and Serpens SMM~1 \citep{casali93}.  It is located in the Serpens Main cluster and is associated with the Bolocam core Ser-Bolo~23.  Ser-emb~6 is the brightest Class~0 protostar in Serpens ($\lbol=28~\lsun$), and has the largest envelope mass ($\menv=20~\msun$) of any embedded protostar in the region. 

The 230~GHz map of Ser-emb~6 shows a complex source with both compact and extended emission.  Approximately 58\% of the single dish flux is recovered by CARMA, with a peak visibility amplitude of 2.3~Jy.  In Paper~I we compared the CARMA visibilities, \textit{Spitzer} IRS spectra, and broadband SED to a grid of detailed radiative transfer models calculated with RADMC \citep{dd04}.  The visibilities were best fit by a massive disk ($\mdisk=2.5~\msun$ when scaled for $d=415$~pc), making up 13\% of the total circumstellar mass and accounting for most of the flux at $uv$-distances $30-150$~k$\lambda$.  
Using only the flux at 50~k$\lambda$ results in a disk mass of $1.7~\msun$, or 9\% of the circumstellar mass, an underestimate of 30\% compared to detailed modeling. 

\citet{choi09} suggests that Ser-emb~6 is a proto-binary source with a separation of 800~AU.
This second source corresponds to the small peak seen to the north-west of Ser-emb~6 (Figure~\ref{mapfig}).  Given its location along the outflow axis, and the clumpy nature of the envelope, we concluded in Paper~I that this emission is associated with the outflow, but \citet{choi09} argues that the source is a second disk component based on the spectral index from 3.5~cm to 7~mm.

\subsection{Ser-emb~7}
Ser-emb~7 is located in Cluster B, and is associated with the Ser-Bolo~8 core.  Two other nearby protostars, Ser-emb~3 ($30\arcsec$ away) and Ser-emb~9 ($26\arcsec$ away) are associated with the same single dish core, but both were only marginally detected in initial 110~GHz CARMA map (Section~\ref{appendsec}), suggesting that most of the core mass is associated with Ser-emb~7.  Another possibility is that Ser-emb~3 and emb~9 have envelopes with a flat central density profile, which is resolved out by the interferometer.  
Neither Ser-emb~3 or emb~9 are included in our 230~GHz study due to the lack of observed millimeter flux. 

The 230~GHz map of Ser-emb~7 shows a single extended source that appears elongated in the north-south direction.  
Only 29\% of the single dish flux is recovered by CARMA.
The flux at 50~k$\lambda$ suggests a disk mass of $0.15~\msun$, or 4\% of the total circum-protostellar mass, but the two nearby protostars may contribute to the single dish flux if they have envelopes that are resolved out by the interferometer.

\subsection{Ser-emb~8}
Ser-emb~8, also known as S68N \citep{mcmullin94} or SMM~9 \citep{casali93}, is located in the Serpens Main cluster.  
Although two Class~I protostars are located within the FWHM of the Bolocam core Ser-Bolo~22, the core is centered on Ser-emb~8 and the majority of envelope mass appears to be associated with this object.

The 230~GHz map indicates a compact source at the position of Ser-emb~8, with two additional sources $\sim 12\arcsec$ and $\sim 20\arcsec$ away.  
Approximately 20\% of the single dish flux is recovered with CARMA; as with Ser-emb~4 and emb~7, nearby sources were likely blended in the Bolocam beam.  
The flux at 50~k$\lambda$ suggests a disk mass of $0.25~\msun$, or 3\% of the total circumstellar mass, likely a lower limit as nearby sources are contributing to the single dish flux.

\subsection{Ser-emb~11 and Ser-emb~17}
Ser-emb~11 and emb~17 are both Class~I sources associated with the Ser-Bolo~14 core, located in Cluster B.  The two protostars are separated by $12\arcsec$, or 5000~AU, suggesting that they may be physically associated.  They have similar SEDs, but Ser-emb~17 is brighter at IRAC wavelengths, resulting in a higher $\tbol$ (117~K compared to 77~K).  In \citet{enoch09a}, the total 1.1~mm flux of Ser-Bolo~14 was divided evenly between them to determine the envelope masses. 

The 230~GHz map reveals two compact sources at the positions of Ser-emb~11 and emb~17.  
Ser-emb~11 is further resolved into two components separated by $2.1\arcsec$ or 870~AU, with a flux ratio of approximately 2.  This is the only source in our sample with clear evidence for a binary or multiple component for separations $<2000$~AU.  

The visibility amplitudes of both Ser-emb~11 and emb~17 have narrow peaks, indicating very extended envelopes, and nearly constant amplitudes to $uv$-distance $>100$~k$\lambda$, suggesting compact disk components.
Together, the peak visibility amplitudes of Ser-emb~11 and emb~17 account for approximately 60\% of the total single dish flux.
The fluxes at 50~k$\lambda$ are similar, resulting in disk masses of $0.13~\msun$ (emb~11) and $0.15~\msun$ (emb~17), together approximately 4\% of the total circumstellar mass.  
Note that visibilities and disk mass estimates are calculated only for the brighter emb~11 component (W).

\subsection{Ser-emb~15}
Ser-emb~15 is a Class~I source located to the east of Cluster B, in the Ser-Bolo~24 core.  The Bolocam core is also associated with the Class~0 source Ser-emb~2, which is $28\arcsec$ or 12000~AU away from emb~15.  Ser-emb~2 was not detected in preliminary 110~GHz CARMA imaging (Figure~\ref{mm3fig}).  
As there is significant 1.1~mm emission associated with emb~2 (see also Figure~\ref{spitzfig}), the non-detection may be due to the lack of a compact disk combined with  a flattened envelope density profile.

The 230~GHz map of Ser-emb~15 shows a single compact source.  The visibility amplitudes are nearly constant for $uv$-distances $5-200$~k$\lambda$, without the peak at small $uv$ distances expected for an extended envelope.  Rather, the visibilities are consistent with a bright unresolved disk.  The CARMA observations recover 54\% of the total single dish flux associated with Ser-emb~15.  
The flux at 50~k$\lambda$ suggests a disk mass of $0.15~\msun$, only 13\% of the total circum-protostellar mass. If a large fraction of the 1~mm flux is coming from the disk, however, the envelope mass calculated assuming $T_D=15$~K is likely overestimated.  Thus, Ser-emb~15 appears to have a compact disk surrounded by a relatively low mass envelope, as expected if Class~I sources have exhausted at least half of their envelope mass \citep[e.g.][]{andre94}.

\section{Multiplicity}\label{multsec}

The very high resolution millimeter maps presented here are also sensitive to binary and multiple sources, assuming that each source has a separate disk or envelope component.  The multiplicity fraction at very early times is essential for understanding the formation of multiple systems, but very little is currently known about multiplicity in the Class~0 stage.

Although our observations do not have uniform sensitivity, the sensitivity to small scales (down to 250~AU) makes this sample a valuable addition to multiplicity studies.
In only one source, Ser-emb~11, do we find strong evidence for multiplicity on scales $<2000$~AU ($5\arcsec$).  Ser-emb~11 (E) and (W) are separated by 870~AU.  Although Ser-emb~11 is a Class~I source, it is very near the Class~0 cutoff ($\tbol=77$~K), and is probably young enough to be included with the Class~0 sample.  
For our sample of 9 sources, this yields a multiplicity ratio of approximately 10\%.   
Ser-emb~6 may also be a binary source with 800~AU separation based on the VLA observations of \citet{choi09}, so it is possible that the fraction is as high as 20\%.

Approximately 21\% of Class I protostars have companions in the $250-2000$~AU range \citep*{connelley08}, with similar binary frequencies in more evolved Class~II ($\sim16$\%; Kraus et al., in preparation) and main sequence solar-type stars ($\sim10$\%; \citealt{rag10}). 
\citet{maury10} recently observed 5 Class~0 protostars with sub-arcsecond resolution with IRAM-PdBI to study the fraction of binary and multiple systems down to 50~AU separations.  No Class~0 binaries were found with separations $<1500$~AU, in contrast to the $\sim20$\% multiplicity fractions observed in more evolved objects.  

Combining our sample with the \citet{maury10} study, the resulting low Class~0 multiplicity fraction ($\sim10$\%) is generally consistent the suggestion of \citet{maury10}, that the binary fraction in the 75-2000~AU range increases from the Class~0 to Class~I stage, possibly due to dynamical evolution of initially very tight ($<75$~AU) or very wide ($>2000$~AU) systems.  With still relatively few sources, however, the statistical uncertainties are too large to conclusively determine that Class~0 sources have fewer intermediate scale binaries than more evolved protostars.

\section{Summary}\label{sumsec}

We have completed high-resolution, high-fidelity millimeter imaging of nine embedded protostars in the Serpens molecular cloud with the CARMA interferometer.  Our sample is based on the complete ($\menv\gtrsim0.25~\msun$, $\lbol\gtrsim0.05~\lsun$) census of embedded protostars in Serpens from \citet{enoch09a}, and includes 6 of the 9 Class~0 sources in Serpens.  

230~GHz continuum observations were completed in B, C, D, and E configurations, with small 7-point mosaics in the compact D and E configurations; the resulting images are sensitive to spatial scales from $10^2$ to $10^4$ AU.  
Our high resolution maps are also sensitive to binaries with separations $>250$~AU.  In only one source is there clear evidence for multiple components within 2000~AU, suggesting a tentative multiplicity fraction of only 10\% for very deeply embedded protostars. 

For un-blended sources, our CARMA observations recover $50-95$\% of the single dish Bolocam flux, assuming a dust spectral index of $\beta=1.5$. 
In all nine observed sources, 230~GHz emission is seen on both large scales (a few $1000$~AU), indicating an extended envelope, and small scales (down to 170~AU), indicating a compact disk component.  
Three of the nine Serpens Class~0 sources were not observed here due to non-detections or marginal detections in preliminary 110~GHz CARMA data, and these sources likely do not have a significant disk.  Thus, we see evidence for circum-protostellar disks in 6/9 Class 0 sources in Serpens.

We make a first estimate of the disk mass using the visibility amplitude at $50$~k$\lambda$, assuming an optically thin, unresolved disk with $T_D=30$~K and a distance to Serpens of 415~pc \citep{dzib10}.  
Resulting disk mass estimates range from $0.04~\msun$ to $1.7~\msun$, with a mean disk mass of $0.2~\msun$ when the three non-detections are included.
Errors in disk masses may be $\pm50$\%, depending on the envelope contribution at $50$~k$\lambda$, the size of the disk, and the presence of optically thick emission.  For a distance to Serpens of $260$~pc, as assumed in our previous work, masses would deccrease by a factor of 2.5. 
Our mean Class~0 disk mass is similar to that of the larger, but less well defined, Class~0 and Class~I sample of \citet{jorg09}.

Our high disk detection rate in a complete sample suggests that circum-protostellar disks are quite common even in the youngest protostars, with approximately 2/3 of embedded protostars in Serpens having detectable ($\mdisk\gtrsim0.02~\msun$) disks by 0.2~Myr.  

Future work comparing the high-resolution millimeter data, as well as spectral information, with radiative transfer models will provide a more robust determination of the disk masses.

\acknowledgments

The authors thank the referee for comments and suggestions that improved this manuscript.  
Support for this work was provided by NASA through the Spitzer Space Telescope Fellowship Program, through a contract issued by the Jet Propulsion Laboratory (JPL), California Institute of Technology, under a contract with NASA.  DPM was supported by NASA through Hubble Fellowship grant \#HST-HF-51259.01
Support for CARMA construction was derived from the states of California, Illinois, and Maryland, the Gordon and Betty Moore Foundation, the Kenneth T. and Eileen L. Norris Foundation, the Associates of the California Institute of Technology, and the National Science Foundation. Ongoing CARMA development and operations are supported by the National Science Foundation under a cooperative agreement, and by the CARMA partner universities.

\appendix
\section{Preliminary 110~GHz data}\label{appendsec}

Preliminary $\nu=110$~GHz ($\lambda=2.7$~mm) observations of the Bolocam 1.1~mm cores Ser-Bolo~8 (associated with mid-infrared identified protostars Ser-emb~3, emb~7, and emb~9) and Ser-Bolo~24 (associated with Ser-emb~2 and emb~15) were completed with CARMA. 
The 6.1-m and 10.4-m antennas were used to obtain 110~GHz continuum observations in the D and E antenna configurations between 2007 April 20 and 2007 September 04.  

Three correlator bands were configured for continuum observations with 468.75 MHz bandwidth, for a total bandwidth of 2.82~GHz.  
1751+096 was observed approximately every 20 minutes for complex gain calibration.  Absolute flux calibration was accomplished using 5 minute observations of MWC~349.  The overall calibration uncertainty is approximately $\pm20\%$, from the reproducibility of the phase calibrator flux on nearby days. 
The passband calibrator, 1751+096, was observed for 15 minutes during each set of observations, and either optical \citep{corder10} or radio pointing was performed every one to three hours.  
Data reduction was completed as described for the 230~GHz data in Section~\ref{obssec}.

\begin{figure*}[!ht]
\includegraphics[width=7.in]{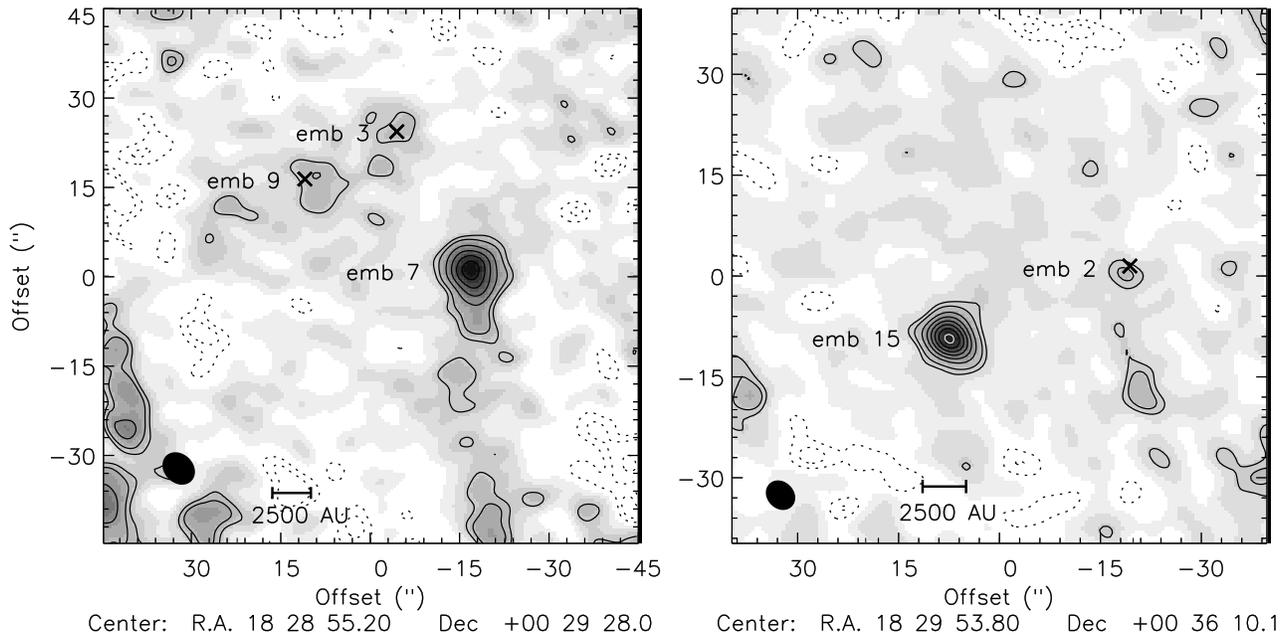}
\caption
{CARMA 110~GHz maps of the Ser-Bolo~8 and Bolo~24 single dish 1.1~mm cores, with which the embedded protostars Ser-emb~2, emb~3, emb~7, emb~9, emb~15 are associated.  Contours are $2,3,5,..15~\sigma$, with lighter contours starting at $15~\sigma$ and negative contours indicated by dotted lines.  The synthesized beam sizes ($5-6\arcsec$) are indicated by solid ellipses.
Ser-emb~2, emb~3, and emb~9 are not detected or only marginally detected at the $3\sigma=4.5$~mJy level, suggesting that very little of the mass traced by the single dish core is associated with these sources.  
\label{mm3fig}}
\end{figure*}

The resulting 110~GHz maps are shown in Figure~\ref{mm3fig}.  The rms noise in both maps is $\sim1.5$~mJy, and the synthesized beam sizes are $5\arcsec-6\arcsec$.  Ser-emb~7 and emb~15 are both strongly detected.  Ser-emb~3 and emb~9 are undetected at the $3\sigma=4.5$~mJy level, although there is some $2\sigma$ emission at the position of both sources.  Ser-emb~2 is marginally detected at the $3\sigma$ level.

Any detected flux in the compact D and E configurations may well include envelope emission, so we adopt an upper limit of 4.5~mJy for the disk flux of all three sources.  We convert this flux upper limit to a disk mass upper limit using equation~\ref{masseq}, where $\kappa_{2.7mm} = 0.003$~cm$^2$~g$^{-1}$ is the dust opacity at 110~GHz.  The opacity is extrapolated from Table~1 column~6 of \citet{oh94} assuming $\kappa \propto \nu^{\beta}$ and $\beta=1.5$.  

This yields a disk mass upper limit of $0.02~\msun$ for Ser-emb~2, emb~3, and emb~9.
While the preliminary 110~GHz maps are less sensitive than the subsequent 230~GHz observations, non-detections place a fairly stringent upper limit on the total envelope mass, and thus on the disk mass.

\clearpage

\end{document}